\documentclass[12pt,a4paper]{article}
\newcommand{\p}{\partial}
\newcommand{\F}{\mathcal{F}}
\newcommand{\VT}{e^{-\frac{1}{4} T^2}}
\newcommand{\intdx}{\int d^{p+1}x}
\newcommand{\nn}{\nonumber\\}
\newcommand{\calO}{\mathcal{O}}
\title{Gauge Fields on Tachyon Matter}
\author{
{\sc Akira Ishida}\footnote{e-mail:ishida@eken.phys.nagoya-u.ac.jp}~
and
{\sc Shozo Uehara}\footnote{e-mail:uehara@eken.phys.nagoya-u.ac.jp}
\vspace{3mm}\\
{\sl Department of Physics Nagoya University}\\
{\sl Chikusa-ku Nagoya 464-8602,Japan}}
\date{}

\begin{document}
\maketitle
\vspace{-87mm}
\begin{flushright}
DPNU-02-17\\ hep-th/0206102\\ June 2002
\end{flushright}
\vspace{55mm}

\maketitle
\begin{abstract}
We study the rolling tachyon including the gauge fields
in boundary string field theory.
We show that there are no plane-wave solutions for the gauge fields
for large time.
The disappearance of the plane-wave solutions indicates
that there are no excitations of the gauge fields
on the tachyon matter, which is consistent with the Sen's conjecture.
\end{abstract}

Recently, Sen \cite{Sen1,Sen2} considered the decay process of
unstable D-branes in bosonic and superstring field theories and
constructed the classical time dependent solutions which describe the
rolling tachyon field toward the bottom of the tachyon potential.
It was shown in \cite{Sen2} that the energy density remains constant
and the pressure approaches zero as the tachyon field rolls toward the
minimum.
This pressure-less gas with nonzero energy density is called the
tachyon matter.
This phenomenon was analyzed by using the Born-Infeld type effective
field theory. It was shown in \cite{Sen3} that there are no plane-wave
solutions for the tachyon field around the minimum of the tachyon
potential and the pressure falls off at late time.
Cosmological considerations for the tachyon matter have been studied
in many papers \cite{cosmology}.

Another analysis was carried out by using the boundary string field
theory (BSFT) \cite{BSFT}--\cite{NP}, which gives some exact results
such as the exact form of the tachyon potential and the exact
solutions of the lower dimensional brane \cite{KMM1,KMM2}.
So it is reasonable to consider the rolling tachyon in the BSFT
framework. Time dependent solution describing the rolling tachyon
was constructed in \cite{ST,Mi}. The solution asymptotically
approaches $T=x^0$ and its behavior for large time describes a
pressure-less gas with nonzero energy density. So, the solution for
large time represents the tachyon matter.
Since the tachyon matter is no longer a D-brane,
it is important to examine whether the excitations of the gauge fields
are absent or not at the bottom of the tachyon potential.
The absence of the open string excitations corresponds to
the absence of the plane-wave solutions in the effective field theory.
In this letter, we consider the equations of motion of the gauge
fields at the bottom of the tachyon potential by using BSFT
and show the absence of the plane-wave solutions.

Let us consider the non-BPS $Dp$-brane in Type II string theory.
Here we restrict ourselves to almost spatially homogeneous time
dependent tachyon field and small fluctuations of the gauge fields.
The BSFT action with the tachyon and the gauge fields is given
by\footnote{We set $\alpha'=2$ and rescale the gauge fields properly
for simplicity. This action is exact as long as $\p_\mu \p_\nu T=0$
and $\p_\rho F^{\mu\nu}=0$ and it is difficult
to examine higher derivative corrections in the BSFT framework.
Here we proceed to investigate with the action eq.(\ref{S0}) and
we will comment on the validity of our result later.}
\cite{TU,KMM2,An}
\begin{equation}
 S=-T_p \intdx\, \VT\, \sqrt{-\det (\eta +F)}~\F
	(G^{\mu\nu} \p_\mu T \p_\nu T)\,,\label{S0}
\end{equation}
where
\begin{eqnarray}
 \F(x)&=&\frac{x\,4^x\, \Gamma(x)^2}{2\Gamma(2x)}
	=1+(2\log2)\,x+\cdots, \\
 G^{\mu\nu}&=&\left( \frac{1}{\eta +F} \right)^{(\mu\nu)}
	= \eta^{\mu\nu}+\calO (F^2) =G^{\nu\mu}, \\
 \det(\eta+F)&=&\det(\eta_{\mu\nu}+F_{\mu\nu})\,,
\end{eqnarray}
and  $T_p$ is the tension of the unstable $Dp$-brane.
Since we consider almost spatially homogeneous tachyon field and small
fluctuation of the gauge field, $\p_i T$ and $F_{\mu\nu}$ (or $A_\mu$)
are small, which we symbolically represent as $\calO(\varphi)$, so
that we have
\begin{eqnarray}
   G^{\mu\nu}\p_\mu T \p_\nu T &=&
	G^{00}\,\dot{T}^2 + \eta^{ij}\p_i T \p_j T
	+\calO(\varphi^3)\nn
  &=&  -\dot{T}^2 + F^{0i}F_{0i}\,\dot{T}^2 + \eta^{ij}\p_i T \p_j T
	+\calO(\varphi^3)\,,\label{eq:GTTexp}
\end{eqnarray}
where $\dot{*}\equiv\p_0 *$ and $i,j=1,\cdots ,p$.

Let us compute the Hamiltonian density.
The conjugate momenta $P$ and $\Pi^{\mu}$ of the tachyon $T$ and the
gauge field $A_\mu$, respectively,  are given by
\begin{eqnarray}
  P= \frac{\delta S}{\delta \dot{T}} &=& -2T_p\,\VT\,
	\sqrt{-\det(\eta+F)}~G^{0\mu}\p_\mu T\,\F'\,,\\
  \Pi^\mu = \frac{\delta S}{\delta \dot{A_\mu}}
  &=& -T_p\,\, \VT \sqrt{-\det(\eta +F)}\nn
  &&\hspace{11ex}
	\times\left[f^{\mu0}\F
	-2\left\{G^{\mu\lambda}f^{\rho0}
        +G^{\rho0}f^{\mu\lambda}\right\}
	\p_\lambda T \p_\rho T\, \F' \right] ,
\end{eqnarray}
where
\begin{equation}
  f^{\mu\nu}= \left(\frac{1}{\eta+F}\right)^{[\mu\nu]} =
	-F^{\mu\nu} + \calO(F^3)= -f^{\nu\mu}.\label{eq:f}
\end{equation}
Then the Hamiltonian density is
\begin{eqnarray}
 \mathcal{H}&=& P \dot{T} +\Pi^{\mu} \dot{A}_\mu -\mathcal{L} \nn
  &=& - T_p\,\, \VT \sqrt{-\det(\eta+F)}
	\,\bigg[ (f^{\mu0}\dot{A}_\mu - 1)\,\F \nn
  &&\hspace{13ex}+ 2\{ G^{0\nu}\p_\nu T\, \dot{T} - \dot{A}_\mu\p_\nu
	T\p_\lambda T(G^{\mu\lambda}f^{\nu0}
	+G^{\nu0}f^{\mu\lambda}) \} \F' \bigg].
\end{eqnarray}

We consider the behavior of the rolling tachyon $T$ at $x^0\to
\infty$ as was discussed in \cite{ST,Mi}.
Here, we set the initial conditions of $T=0$ and $\dot{T}=+0$ for
simplicity.
It was shown that the tachyon, which is rolling down to the bottom of
the tachyon potential, never stops~\cite{ST}.
Thus, $T$ becomes infinity and hence $e^{-T^2/4}\to 0$ as
$x^0\to\infty$.
Since $\mathcal{H}$ should be finite, the rolling
tachyon must hit a singularity and $\F$ should become infinity as
$x^0\to\infty$.

$\F(z)$ and $\F'(z)$ have singularities at $z=-n$\quad
($n=1,2,\cdots$) and the nearest singular point from
$z=0$\footnote{Note that $z=0$ corresponds to $\dot{T}=0$.} is
$z=-1$, and hence we require the following condition,
\begin{equation}
 \label{limit1}
	G^{\mu\nu}\p_\mu T\p_\nu T\to -1\,.
\end{equation}
The asymptotic behavior of $\F(z)$ and $\F'(z)$
near $z=-1$ are as follows,
\begin{equation}
 \F(z) \sim \frac{-1}{2(z+1)}\,, \hspace{3ex}
	\F'(z) \sim \frac{1}{2(z+1)^2}\,.\label{Fasym}
\end{equation}
Since $\F'(z)$ is more singular than $\F(z)$ at $z=-1$,
$e^{-T^2/4}\,\F'$ should be finite as $x^0 \to \infty$ because of
energy conservation. From eq.(\ref{limit1}), we write the asymptotic
equation for $\dot{T}$ as (cf. eq.(\ref{eq:GTTexp}))
\begin{equation}
  \dot{T} \sim \frac{1}{\sqrt{-G^{00}}}\,\left(
	1 +\frac{1}{2}\eta^{ij}\p_i T \p_j T
	+ \calO(\varphi^3)+ \epsilon(x)\right), \label{dT}
\end{equation}
where $\epsilon(x)$ is a small perturbation.\footnote{We have assumed
that the BSFT action is valid for this form of $T$ as is mentioned in
the previous footnote. We note that a reliable result has been
obtained in the similar situation \cite{ST}.}
Note that we can see from eq.(\ref{dT}) that the tachyon field becomes
infinity as $x^0\to\infty$. Furthermore, eqs.(\ref{Fasym}) and
(\ref{dT}) lead to the asymptotic form of $\F'$,
\begin{equation}
 \F' \left( G^{\mu\nu}\p_\mu T\p_\nu T\right)
 \sim \frac{1}{8 \epsilon(x)^2} (1+\calO (\varphi^2))\, .
\end{equation}
Due to the requirement that $e^{-T^2/4}\F'$ should be
finite, we can determine $\epsilon(x)$ as
\begin{equation}
 \epsilon(x)\sim -C(x)\, \exp \left(-\frac{1}{8}\, T^2 \right) \,,
 \hspace{3ex} C(x)=C+\calO (\varphi^2)\,,
\end{equation}
where $C$ is a constant to be determined by energy conservation.
Thus, in the $x^0 \to \infty$ limit, we have
\begin{equation}
 \label{limit2}
 \VT \F \left(G^{\mu\nu}\p_\mu T\p_\nu T\right) \to 0,~
 \VT \F'\left(G^{\mu\nu}\p_\mu T\p_\nu T\right) \to
	\frac{1}{8 C^2}+\calO(\varphi^2).
\end{equation}

Now we discuss the equations of motion of the gauge fields both at
$x^0=0$ and $x^0 \to \infty$. They are derived from eq.(\ref{S0}),
\begin{equation}
\p_\nu \left[ \VT \sqrt{-\det(\eta+F)}\,\left\{
	f^{\mu\nu}\F - 2 \left( G^{\mu\rho} f^{\lambda\nu} +
	G^{\nu\lambda} f^{\mu\rho}\right) \p_\lambda T\p_\rho T
	\,\F'\right\}\right]=0. \label{eom1}
\end{equation}
First, we consider the case where the tachyon is on the top of the
potential. The equations of motion of the gauge fields at $x^0=0$ can
be obtained by plugging the initial conditions, $T=0$, $\dot{T}=+0$,
into eq.(\ref{eom1}). Since we consider small fluctuations of the
gauge fields, we ignore $\calO(\varphi^2)$ terms and hence the
second term in the brace does not contribute in this case.
Using eq.(\ref{eq:f}), we have the ordinary Maxwell equation,
\begin{equation}
	\p_\mu F^{\mu\nu}=0\,.
\end{equation}
Then it becomes, in the Coulomb gauge,\footnote{We can put $A^0=0$,
as usual.}
\begin{equation}
	\p_\mu \p^\mu A^i =0\,.
\end{equation}
Plugging a plane-wave solution,
\begin{equation}
	A^i=a^i\, e^{ik_\mu x^\mu}\,, \label{pwave}
\end{equation}
into this equation, we get
\begin{equation}
	k_\mu k^\mu =0\,.
\end{equation}
Therefore, there exist the plane-wave solutions at $x^0=0$.

Next, we consider the gauge fields at $x^0 \to \infty$.
Similarly, we ignore $\calO(\varphi^2)$ terms and hence, for
example, only $\dot{T}^2$ in $\p_\lambda T \p_\rho T$ contributes.
{}From eqs.(\ref{eq:f}), (\ref{dT}) and (\ref{limit2}), we obtain
the equations of motion at $x^0 \to \infty$,
\begin{equation}
  \p_0 F^{0 k}=0\,,\hspace{3ex} \p_i F^{i0}=0\,,
\end{equation}
where $k=1,\cdots ,p$.
Then, at $x^0 \to \infty$ we have
\begin{equation}
	\p_0 \p^0 A^k =0\,. \label{eom3}
\end{equation}
Plugging (\ref{pwave}) into eqs.(\ref{eom3}), we get
\begin{equation}
	k^0 =0\,.
\end{equation}
Thus, contrary to the $x^0=0$ case, the plane-wave solution is absent
for large time. From the above analysis, we conclude that the
excitations of the gauge fields on a brane disappear as the tachyon
field evolves toward the minimum of the potential, even though the
energy density is conserved. This is consistent with the Sen's
conjecture in which the brane will disappear at the minimum of the
potential. One comment is in order: One may think that the
result will be altered if the action (\ref{S0}) has higher derivative
terms. However, such terms as can change eq.(\ref{eom3}) for a
plane-wave equation will take the form of $f(\p^2 T,\p^3 T,\dots)\cdot
F^2$ where $f$ is a function satisfying $f(0)=0$, and since $\dot T\to
1 +\calO(\varphi^2)$ and hence $f(\p^2 T,\p^3 T,\dots)\to 0$ at
$x_0\to\infty$, we can expect that existence of those terms does not
alter our result.\footnote{$\p F,\p^2 F,\dots$ terms in the action
will not make the transverse modes of the gauge fields.}

Although we have focused on the decay process of an unstable D-brane,
it will be interesting to extend our analysis to brane-anti-brane
systems to see pair annihilation processes of D-branes.
It will be also interesting to consider the space dependent rolling
tachyon in the decay process of a non-BPS brane into a lower
dimensional brane.

\vspace{5mm}
{\bf Acknowledgments:}
The work of SU is supported in part by the Grant-in-Aid for Scientific
Research No.13135212.


\end{document}